\documentclass[a4paper,fleqn]{cas-dc}

\usepackage[numbers,sort&compress]{natbib}

\usepackage[version=4]{mhchem}

\def\tsc#1{\csdef{#1}{\textsc{\lowercase{#1}}\xspace}}
\tsc{WGM}
\tsc{QE}
\tsc{EP}
\tsc{PMS}
\tsc{BEC}
\tsc{DE}

\begin{document}
\let\WriteBookmarks\relax
\def\floatpagepagefraction{1}
\def\textpagefraction{.001}

\shorttitle{From Surface Reactions to Inhibitor Coverage: A Machine-Learning-Assisted Computational Study of AS-ALD}

\shortauthors{G. Kim et~al.}

\title [mode = title]{From Surface Reactions to Inhibitor Coverage: A Machine-Learning-Assisted Computational Study of Area-Selective Atomic Layer Deposition}                      

\author[1]{Gijin Kim}[orcid=0009-0004-5787-2659]
\credit{Conceptualization, Methodology, Software, Validation, Formal analysis, Investigation, Data curation, Writing -- original draft, Visualization}

\author[1,2,3]{Seungwu Han}[orcid=0000-0003-3958-0922]
\cormark[1]
\ead{hansw@snu.ac.kr}
\credit{Conceptualization, Resources, Supervision, Project administration, Funding acquisition}

\author[4]{Youngho Kang}[orcid=0000-0003-4532-0027]
\cormark[1]
\ead{youngho84@inu.ac.kr}
\credit{Conceptualization, Supervision, Writing -- review \& editing}

\affiliation[1]{organization={Department of Materials Science and Engineering, Seoul National University},
    city={Seoul},
    postcode={08826},
    country={Republic of Korea}}

\affiliation[2]{organization={AI Center, Korea Institute for Advanced Study},
    city={Seoul},
    postcode={02455},
    country={Republic of Korea}}

\affiliation[3]{organization={Research Institute of Advanced Materials, Seoul National University},
    city={Seoul},
    postcode={08826},
    country={Republic of Korea}}

\affiliation[4]{organization={Department of Materials Science and Engineering, Incheon National University},
    city={Incheon},
    postcode={22012},
    country={Republic of Korea}}

\cortext[cor1]{Corresponding author}

\begin{abstract}
Area-selective atomic layer deposition (AS-ALD) enables selective material growth for the fabrication of nanoscale semiconductor devices. Predicting inhibitor coverage is essential for understanding and optimizing surface blocking in AS-ALD. Here, we present a computational workflow combining a pretrained universal machine-learning interatomic potential with off-lattice grand canonical Monte Carlo simulations and apply it to inhibitor adsorption and precursor blocking on amorphous \ce{SiO2}. Realistic surface models reveal substantial site-dependent variations in reaction energies and activation barriers, with silanol groups serving as the primary adsorption sites for aminosilane and related inhibitors and metal precursors. Coverage simulations show that inhibitors producing identical adsorbed fragments can yield different coverages because the intact molecules impose distinct steric constraints, leading to different accessible site densities and blocking capabilities. Beyond coverage, we highlight the importance of the reactivity of adsorbed inhibitor fragments toward precursors in determining inhibitor stability and blocking performance. Together, these findings demonstrate the need to consider both steric effects and chemical reactivity on realistic amorphous surfaces. Our calculations connect local reaction chemistry with inhibitor coverage and provide a computational basis for understanding inhibition mechanisms and selecting suitable inhibitor--precursor combinations for AS-ALD.
\end{abstract}

\begin{graphicalabstract}
\includegraphics{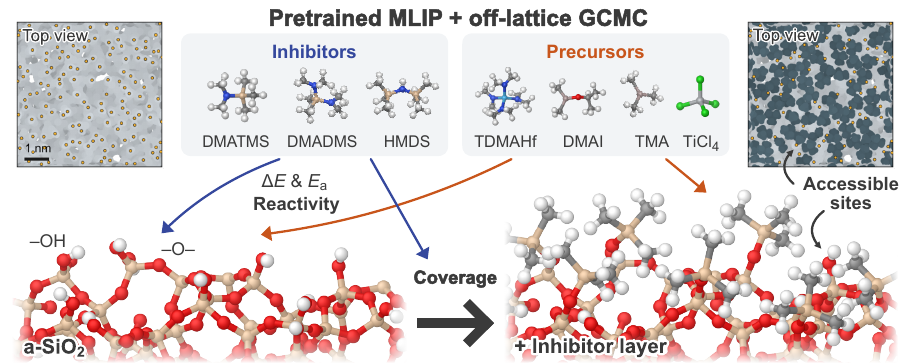}
\end{graphicalabstract}

\begin{highlights}
\item A pretrained MLIP and off-lattice GCMC link surface reactions to inhibitor coverage
\item Barriers on amorphous \ce{SiO2} depend on the local site as well as on the molecule
\item Inhibitors leaving the same fragment reach different coverages by steric packing
\item Sites left open by each inhibitor follow the blocking order observed in experiment
\item TMA can modify the adsorbed trimethylsilyl fragment, whereas \ce{TiCl4} leaves it intact
\end{highlights}


\maketitle

\section{Introduction}

Area-selective atomic layer deposition (AS-ALD) has attracted growing interest as a bottom-up approach to semiconductor fabrication that enables self-aligned deposition on designated growth areas~\cite{Mackus.2019,Parsons.2020,Cao.2020,Yasmeen.2022,Liu.2024dc5,Rhee.2025cpw}. By exploiting differences in surface chemistry to control where material grows, AS-ALD can reduce reliance on repeated patterning and etching steps and help mitigate the limitations of top-down fabrication. These limitations become increasingly important with continued semiconductor device scaling, as overlay errors and dimensional variations introduced during patterning and etching contribute to edge placement error, a major challenge to further scaling~\cite{Raley.2019,Clark.2018}.

Selective deposition requires growth and non-growth surfaces to respond differently to the same ALD chemistry. Intrinsic differences in surface reactivity can provide selectivity for certain material combinations, but the resulting nucleation delay on the non-growth area is often limited~\cite{Liu.2024dc5,Yarbrough.2021}. Inhibitors offer a means of suppressing growth on the non-growth surface across a broader range of material combinations, with blocking effectiveness adjustable through molecular design~\cite{Lee.2025mh,Liu.2024dc5}. Common approaches employ self-assembled monolayers (SAMs) or small-molecule inhibitors (SMIs). SMIs offer practical advantages for AS-ALD in advanced semiconductor fabrication: their small molecular size and compatibility with vapor-phase delivery facilitate integration into ALD processes, including repeated inhibitor dosing during successive cycles~\cite{Mameli.2017,Yarbrough.2021,Liu.2024dc5}. In comparison, assembling densely packed SAMs from long-chain molecules can require prolonged surface treatment~\cite{Liu.2024dc5}.

Experimental studies have demonstrated that inhibitor adsorption can saturate before all reactive surface groups are consumed because steric hindrance limits further uptake~\cite{Soethoudt.2019,Clerix.2023}. Some residual sites remain accessible to ALD precursors~\cite{Merkx.2022}; for example, hydroxyl groups remaining on inhibitor-treated silica have been observed to react during subsequent precursor exposure~\cite{Clerix.2023}. Furthermore, inhibitors that produce the same surface termination can exhibit markedly different blocking effectiveness~\cite{Dongen.2023}, and sequential dosing of different inhibitors can improve growth suppression~\cite{Xu.202223f,Dongen.2023}. These observations highlight the importance of inhibitor coverage and precursor accessibility in determining the effectiveness of surface passivation for the non-growth area.

AS-ALD chemistry, including inhibitor adsorption and precursor reactions, has been widely investigated using density functional theory (DFT) simulations, providing valuable insights into local reaction mechanisms~\cite{Pieck.2025,Khan.2018,Shearer.2025qh2,Wellmann.2024,Kim.2026}. However, these studies have generally focused on reactions at individual surface sites, providing limited information about inhibitor packing and steric constraints on coverage~\cite{Kim.2024hgs,Pieck.2025}. They have also often employed crystalline substrate models, which do not fully capture the structural heterogeneity of the amorphous substrates widely used in experiments~\cite{Yarbrough.2022,Pieck.2025}. Together, these limitations restrict our understanding of inhibitor coverage and its relationship to precursor accessibility and blocking effectiveness.

Machine-learning interatomic potentials (MLIPs) enable efficient atomistic simulations by predicting energies and forces at substantially lower computational cost than DFT~\cite{Focassio.2025,Deringer.2016,Erhard.2022,Nakata.2022}. In particular, pretrained universal models such as SevenNet-omni~\cite{Kim.2026omni}, MACE-MH-1~\cite{Batatia.2025mh}, and UMA~\cite{Wood.2025} allow simulations across diverse chemical systems without constructing a system-specific potential from scratch. Recent applications have demonstrated their utility in investigating atomic structures, reaction pathways, and finite-temperature properties~\cite{Wander.2025,Kim.2024dmai,Kim.2026pecvd,DellaPia.2025,Ju.2025,An.2025,Cho.2026igzo,Moon.2026}. These advances provide opportunities to study larger systems and sample a broader range of configurations, facilitating the investigation of heterogeneous surfaces and coverage-dependent adsorption.

Beyond an efficient interatomic potential, determining inhibitor coverage requires a sampling method that captures both molecular packing and reaction energetics. Previous approaches, such as random sequential adsorption, have described steric constraints on inhibitor uptake without explicitly accounting for reaction energetics~\cite{Li.2022,Tezsevin.2023,Yu.2024ma,Khan.2018,Gu.2024,Merkx.2020}. Monte Carlo methods incorporating reaction energetics have also been employed, but have typically relied on on-lattice models with predefined event tables~\cite{Yun.2022v66,Wang.2024e6a}. Such representations limit the description of irregular reactive-site distributions and the configurational freedom of adsorbed molecules. Coverage studies that combine realistic amorphous surfaces, explicit molecular structures, and configuration-dependent reaction energetics therefore remain limited.

In this work, we present a computational workflow combining a pretrained universal MLIP with off-lattice grand canonical Monte Carlo (GCMC) simulations. To demonstrate its predictive capability, we investigate aminosilane and related inhibitors for suppressing the adsorption of Al, Ti, and Hf precursors on amorphous \ce{SiO2}, which serves as the non-growth surface. We first determine the surface silanol density under experimentally relevant conditions and examine inhibitor and precursor adsorption pathways, validating the reaction energetics against DFT calculations. We then perform GCMC simulations on a $5.92 \times 5.92$~nm$^2$ substrate, using on-the-fly reaction energies and a steric filter to account for molecular overlap, to evaluate inhibitor coverage, precursor accessibility, and the effects of sequential inhibitor dosing. We also assess precursor reactions with adsorbed inhibitor fragments. Our findings highlight the importance of steric constraints on inhibitor uptake, alongside the chemical properties of the anchored fragments, in determining the blocking effectiveness of inhibitor layers. Sequential dosing further reduces precursor accessibility, consistent with experimentally observed improvements in growth inhibition.

\section{Computational methods}
\label{sec:Methods}

\subsection{Machine-learning interatomic potential}
We use SevenNet-omni~\cite{Kim.2026omni}, a pretrained universal MLIP based on an E(3)-equivariant message-passing graph neural network, as implemented in the SevenNet package~\cite{Park.2024sevennet} with FlashTP acceleration~\cite{Lee.2025flashtp}. We employ its mpa channel, which corresponds to the PBE level of DFT. Its energy and force predictions for \ce{SiO2}, inhibitor molecules, and precursor molecules are validated against DFT calculations, as shown in Figure~S3. Because dispersion interactions contribute to molecular adsorption on \ce{SiO2}, we account for them using Grimme's D3 dispersion correction with Becke--Johnson damping~\cite{Grimme.2010,Grimme.2011}. Molecular dynamics (MD) simulations are performed using LAMMPS~\cite{Thompson.2022}, whereas structural relaxations, nudged elastic band (NEB) calculations, and GCMC simulations are carried out using the Atomic Simulation Environment (ASE)~\cite{Larsen.2017}.

\subsection{DFT calculations}
\label{sec:dft}
DFT calculations are performed to provide reference data for MLIP validation using the Vienna \textit{Ab initio} Simulation Package (VASP) with the projector augmented-wave (PAW) method~\cite{Kresse.1996,Kresse.19962o} and the Perdew--Burke--Ernzerhof (PBE) exchange-correlation functional~\cite{Perdew.1996,Perdew.1996ifi}. We use a plane-wave energy cutoff of 480~eV, $\Gamma$-point-only \textit{k}-point sampling, Gaussian smearing with a width of 0.05~eV, and an electronic convergence criterion of $10^{-4}$~eV. The same D3 dispersion correction with Becke--Johnson damping used in the MLIP calculations is applied. A dipole correction is included along the $z$ direction to account for spurious electrostatic interactions arising from the net dipole moment of the asymmetric amorphous slab under periodic boundary conditions.

\subsection{Simulation protocols}

Amorphous \ce{SiO2} (a-\ce{SiO2}) structures are generated using MLIP-based melt--quench MD simulations in the NVT ensemble, with temperature controlled by a Nos\'e--Hoover thermostat~\cite{Hoover.1985}. The initial cells are generated by randomly distributing Si and O atoms at the experimental mass density of 2.2~g/cm$^3$. Each cell is first premelted at 6000~K for 10~ps using a 1-fs time step, then melted at 4000~K for 200~ps and quenched from 4000~K to 300~K over 400~ps, corresponding to a cooling rate of 9.25~K/ps. A 2-fs time step is used for both the melting and quenching stages. The quenched structures are subsequently relaxed using the FIRE optimizer until the maximum force falls below 0.03~eV/\AA{}. The resulting a-\ce{SiO2} structures agree with the DFT structures of a previous work~\cite{Kim.2026pecvd} (Figure~S1).

\begin{figure}[t]
\centering
\includegraphics[width=\columnwidth]{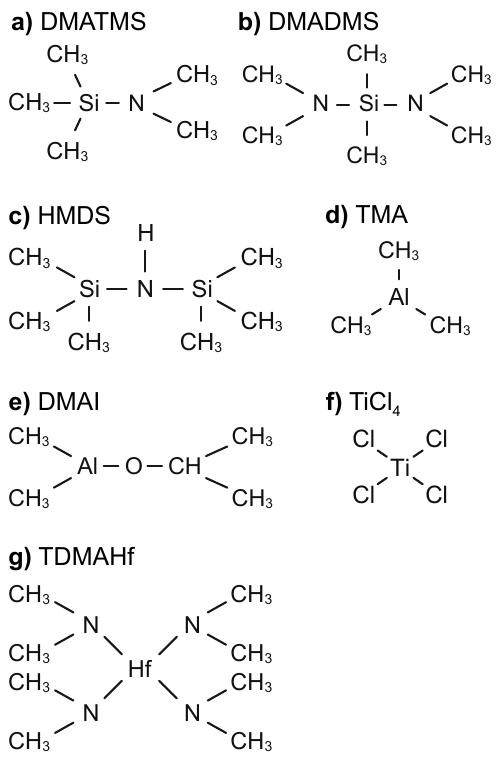}
\caption{Inhibitors and precursors considered in this work. (a) (N,N-dimethylamino)trimethylsilane (DMATMS), (b) bis(N,N-dimethylamino)dimethylsilane (DMADMS), and (c) hexamethyldisilazane (HMDS) are the inhibitors; (d) trimethylaluminum (TMA), (e) dimethylaluminum isopropoxide (DMAI), (f) \ce{TiCl4}, and (g) tetrakis(dimethylamido)hafnium (TDMAHf) are the precursors.}
\label{fig:molecules}
\end{figure}

{\sloppy As inhibitors, we consider the well-known aminosilanes (N,N-di\-methyl\-amino)\-tri\-methyl\-silane (DMATMS) and bis(N,N-di\-methyl\-amino)\-di\-methyl\-silane (DMADMS), as well as the silazane hexa\-methyl\-di\-silazane (HMDS)~\cite{Khan.2018,Soethoudt.2020,Xu.202223f,Dongen.2023}. Their molecular structures are shown in Figure~\ref{fig:molecules}. As ALD precursors, we focus on tri\-methyl\-aluminum (TMA) and di\-methyl\-aluminum iso\-propoxide (DMAI) for Al deposition, \ce{TiCl4} for Ti deposition, and tetrakis(di\-methyl\-amido)\-hafnium (TDMAHf) for Hf deposition. These precursors span a range of ALD chemistries involving methyl, alkoxide, chloride, and amido ligands.\par}

Activation energies are determined using the climbing-image nudged elastic band (CI-NEB) method~\cite{Henkelman.2000,Henkelman.2000dvp}, as implemented in ASE. Depending on the reaction pathway, five to nine intermediate images are used with a spring constant of 0.5~eV/\AA$^2$. The images are optimized using the fast inertial relaxation engine (FIRE) algorithm~\cite{Bitzek.2006}. Structural relaxations are converged to a maximum force below 0.03~eV/\AA{}, whereas CI-NEB calculations are converged to a maximum NEB force below 0.05~eV/\AA{}. DFT activation barriers are determined using the CI-NEB method as implemented in the VTST extension to VASP.

Inhibitor coverages are calculated using off-lattice GCMC simulations. For each trial move, an active site is randomly selected on the surface. Trial moves for inhibitor adsorption through ligand exchange with surface silanol groups (see Section~\ref{sec:pathways} for detailed mechanisms) or the corresponding reverse reaction are accepted or rejected using criteria that incorporate reaction energies evaluated after structural relaxation to a maximum force below 0.05~eV/\AA{}. The theoretical basis for these acceptance rules is detailed in Supplementary Note~S5. The simulations are performed on a 2592-atom a-\ce{SiO2} slab at $T = 423$~K, with atoms in the bottom region held fixed to preserve the bulk structure. The inhibitor partial pressures, which enter the corresponding chemical potentials, are set to 5~Torr for DMATMS and DMADMS and 35~Torr for HMDS, reflecting the experimental conditions~\cite{Xu.202223f,Dongen.2023,Soethoudt.2019}. All byproducts are assumed to have a partial pressure of 5~Torr.

To account for steric hindrance imposed by previously adsorbed inhibitors, a steric filter is applied before evaluating the acceptance of each reaction move. A trial adsorption move is rejected if any interatomic distance between the intact incoming inhibitor and a neighboring adsorbed inhibitor is less than or equal to the sum of the van der Waals radii~\cite{Alvarez.2013} of the corresponding atoms. The same filter is applied to the reverse reaction, in which a byproduct reacts with an adsorbed inhibitor to regenerate the intact molecule; the regenerated inhibitor is checked for steric overlap with neighboring adsorbed inhibitors. PACKMOL~\cite{Martinez.2009} is used to generate trial configurations with diverse molecular orientations by positioning intact inhibitor molecules and adsorbed inhibitor species at the selected active sites for GCMC sampling and steric filtering. We also include exchange moves between adsorbates and surface \ce{-OH} groups to sample configurational rearrangements associated with surface diffusion and structural reorganization.

\section{Results and discussion}
\subsection{Silanol density on a-\ce{SiO2} surface}
\label{sec:Modeling surfaces with MLIPs}

\begin{figure}[t]
\centering
\includegraphics[width=\columnwidth]{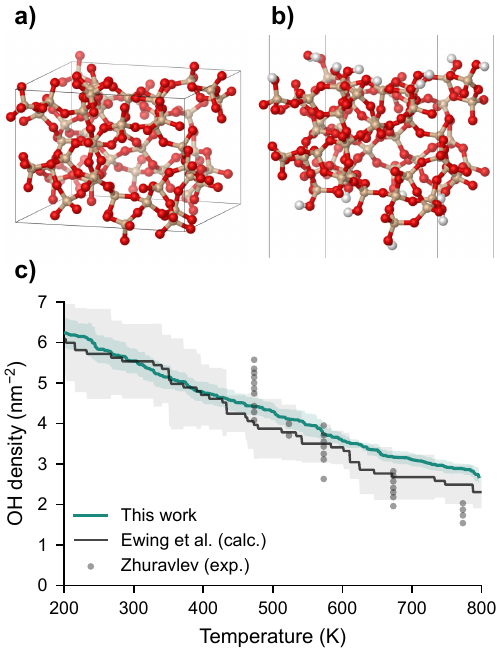}
\caption{Generation of amorphous \ce{SiO2} surfaces. (a) Bulk amorphous \ce{SiO2} cell of 162 atoms generated by melt--quench molecular dynamics. (b) Corresponding slab after cleaving, hydroxylation, and structural relaxation. Si, O, and H atoms are shown in beige, red, and white, respectively. (c) Silanol density of the 2592-atom surfaces as a function of temperature at a water partial pressure of $10^{-6}$~bar. The line represents the average over 12 surfaces, and the shaded band indicates their variation. The calculated values of Ewing et al.~\cite{Ewing.2014} and the experimental values of Zhuravlev~\cite{Zhuravlev.2000} are shown for comparison. The structures in (a) and (b) are rendered with VESTA~\cite{Momma.2011}.}
\label{fig:dehydroxylation}
\end{figure}

Slab models representing a-\ce{SiO2} surfaces are constructed by cleaving the bulk structures and introducing a vacuum region of $\sim$20~\AA{}. We consider two bulk system sizes containing 162 and 2592 atoms, corresponding to 54 and 864 \ce{SiO2} formula units, respectively, and generate 20 and 6 distinct slab models for the two sizes. Each slab exposes a top and a bottom surface, yielding 40 and 12 surfaces, respectively. Fully hydroxylated surfaces are initially prepared by passivating undercoordinated Si and O atoms exposed at both surfaces with hydroxyl (\ce{-OH}) groups and H atoms, respectively [Figure~\ref{fig:dehydroxylation}(b)]. 

The density of surface silanol groups, which serve as potential reaction sites on a-\ce{SiO2}, is sensitive to temperature and water partial pressure. Increasing the temperature or decreasing the water partial pressure promotes dehydroxylation through condensation of neighboring silanol groups, forming a siloxane bridge (\ce{Si-O-Si}) and releasing \ce{H2O}~\cite{Zhuravlev.2000}:
\begin{equation}
\ce{$\equiv$Si-OH + HO-Si$\equiv$ -> $\equiv$Si-O-Si$\equiv$ + H2O(g)} .
\end{equation}
To account for this dependence, we determine the silanol density through free-energy calculations following the procedure of Ewing et al.~\cite{Ewing.2014}, as detailed in Supplementary Note~S2.

Figure~\ref{fig:dehydroxylation}(c) shows the silanol density obtained from the 2592-atom models as a function of temperature at a water partial pressure of $10^{-6}$~bar. The results are largely consistent with previous DFT-based predictions~\cite{Ewing.2014}, and the 162-atom model exhibits a comparable trend (Figure~S2). Both computational predictions reasonably reproduce the experimental data, although they slightly overestimate the silanol density above 600~K. This discrepancy may arise from the limited sampling of proton diffusion and structural reorganization at elevated temperatures~\cite{Zhuravlev.2000,Comas-Vives.2016}. In AS-ALD experiments, inhibitor dosing temperatures typically range from 135 to 250~$^\circ$C~\cite{Khan.2018,Soethoudt.2019,Soethoudt.2020,Dongen.2023}. For the inhibitor adsorption calculations below, we use surface models prepared at 523~K (250~$^\circ$C), with silanol densities of 4.11 and 3.86~OH/nm$^2$ for the 2592- and 162-atom systems, respectively.

\subsection{Inhibitor and precursor adsorption on a-\ce{SiO2} surface}
\label{sec:pathways}

\begin{figure}[t]
\centering
\includegraphics[width=\columnwidth]{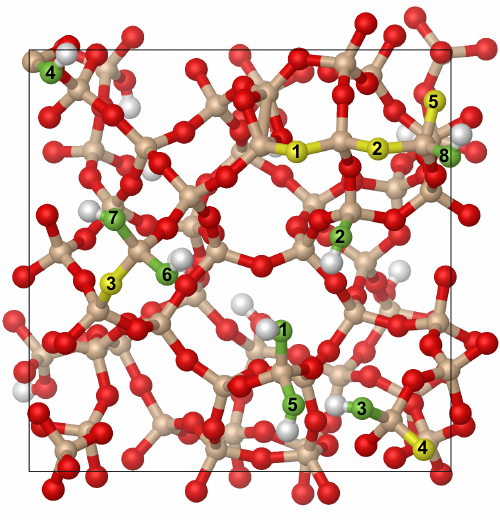}
\caption{Top view of the a-\ce{SiO2} surface used for the reactivity calculations, with sampled oxygen sites numbered. The eight silanol oxygen atoms on the top surface are highlighted in green, and the five highest-lying siloxane bridge oxygen atoms in yellow. Sites within each group are numbered in order of increasing mean activation energy for inhibitor adsorption (see the main text). Si, O, and H atoms are shown in beige, red, and white, respectively.}
\label{fig:sites}
\end{figure}

\begin{figure*}[t]
\centering
\includegraphics[width=\textwidth]{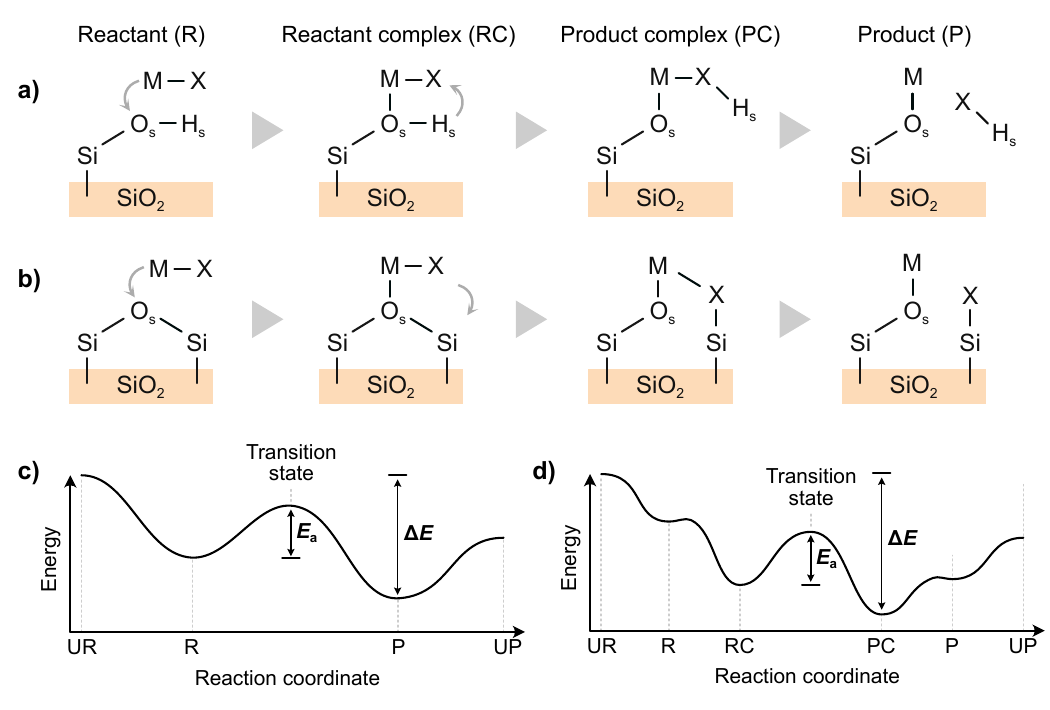}
\caption{Reaction states and energy profiles used in this work. Ligand exchange of a generic inhibitor or precursor M--X, where X denotes the reacting ligand, at (a) a silanol and (b) a siloxane bridge on a-\ce{SiO2}. R denotes the physisorbed reactant. In RC, M bonds to surface oxygen O$_\mathrm{s}$ while the original bonds remain intact. In PC, X bonds to H$_\mathrm{s}$ (silanol) or the surface Si (siloxane) while remaining bound to M. Cleavage of M--X yields P. (c) Single-step energy profile from the gas-phase unbound reactant UR through R and a transition state to P, followed by byproduct desorption to form UP. (d) Multistep profile involving RC and PC. The reaction energy is defined as $\Delta E = E(\mathrm{P})-E(\mathrm{UR})$, or $E(\mathrm{PC})-E(\mathrm{UR})$ when byproduct release from PC is endothermic. The activation energy $E_\mathrm{a}$ is the largest barrier among the steps, each measured from the preceding minimum.}
\label{fig:mechanism}
\label{fig:method}
\end{figure*}

We first investigate the adsorption mechanisms of individual molecules at active sites on the a-\ce{SiO2} surface using MLIP calculations with a 162-atom slab model (Figure~\ref{fig:sites}). This smaller model facilitates validation of MLIP results against DFT calculations, while the adsorption energetics show little dependence on supercell size (Table~S1). To capture the structural heterogeneity of the amorphous surface, we select eight silanol groups and five siloxane bridges as potential reaction sites, sampling more than three molecular orientations at each site. We examine two adsorption reaction pathways: ligand exchange at silanol groups and dissociation at siloxane bridges [Figure~\ref{fig:mechanism}(a) and (b)]. In the case of inhibitors, each adsorption pathway comprises four states: an unbound reactant (UR), a physisorbed reactant (R), a chemically adsorbed product accompanied by a physisorbed volatile byproduct (P), and a final state with the byproduct desorbed (UP) (Figure~\ref{fig:method}). The reaction energy is defined as $\Delta E = E(\mathrm{P}) - E(\mathrm{UR})$. Because physisorption is barrierless, the activation energy is evaluated relative to the physisorbed reactant as $E_\mathrm{a} = E(\mathrm{TS}) - E(\mathrm{R})$, where TS denotes the transition state along the pathway from R to P [Figure~\ref{fig:mechanism}(c)]. On the other hand, metal precursors can form coordinate bonds due to the strong Lewis acidity of the central metal atom. Their adsorption pathways may therefore involve multiple steps, with chemically bound reactant complexes (RC) and product complexes (PC) as intermediates. For such pathways, we report the largest activation barrier among the individual steps [Figure~\ref{fig:mechanism}(d)]. The validation results against DFT calculations are presented in Figure~S4, with mean absolute errors of 0.36 and 0.14~eV for $\Delta E$ and $E_\mathrm{a}$, respectively. Further improvements in accuracy may be achieved through model fine-tuning~\cite{Kim.2026omni,Oh.2026,Radova.2025}. In this work, however, we use the pretrained model directly, as its accuracy is sufficient to reasonably describe reaction energetics and inhibitor coverages, as discussed below. 

\begin{figure*}[t]
\centering
\includegraphics[width=\textwidth]{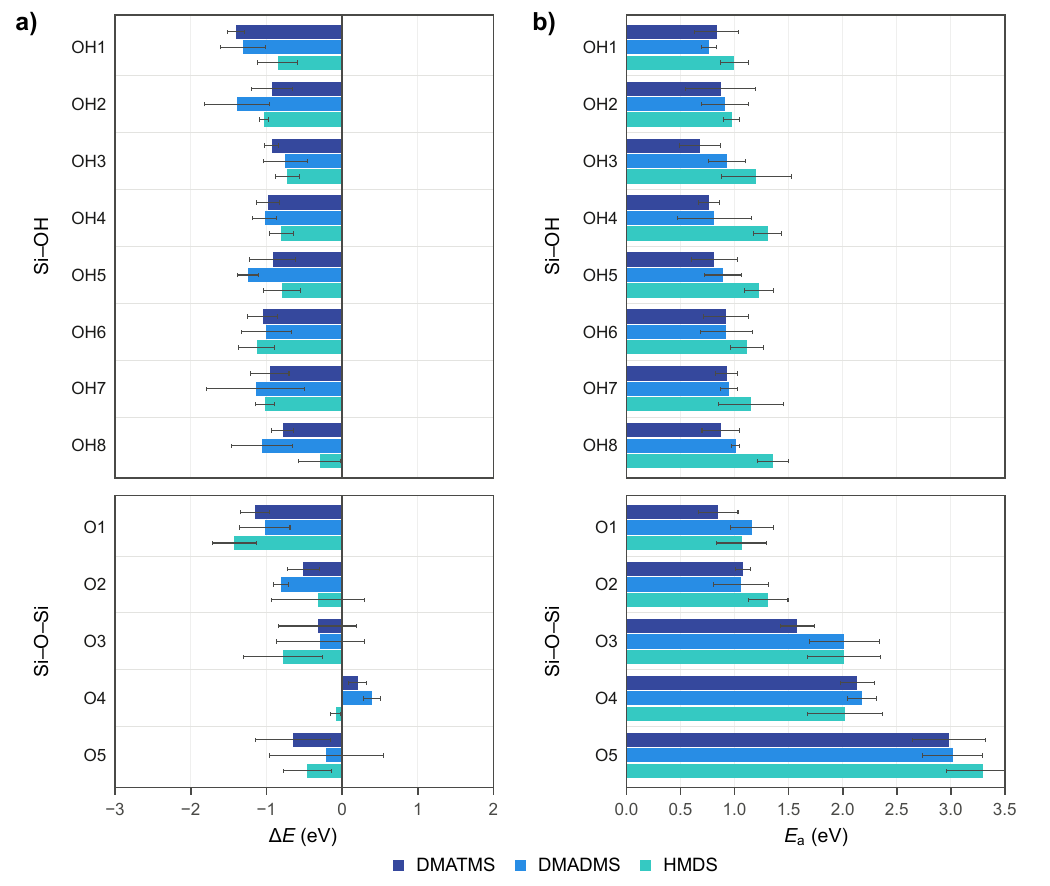}
\caption{(a) Reaction energies $\Delta E$ and (b) activation energies $E_\mathrm{a}$ of DMATMS, DMADMS, and HMDS at the sampled silanol (Si--OH) and siloxane (Si--O--Si) sites. Each row corresponds to a surface site numbered as in Figure~\ref{fig:sites}, and colors distinguish the inhibitors. Bars represent means over three to five independent samples for each inhibitor--site pair, with error bars indicating standard deviations. Within each site type, rows are ordered from top to bottom by increasing activation energy averaged over the three inhibitors. The same site ordering is used in both panels to facilitate comparison.}
\label{fig:reactivity-inh}
\end{figure*}

\subsubsection{Inhibitor adsorption}
\label{sec:path-surf-inh}

All three inhibitors react in a single step from R to P at both site types, without forming RC or PC intermediates, as expected from their tetravalent Si centers. During adsorption at a silanol group, the silanol proton is transferred to the inhibitor nitrogen as the Si--N bond breaks and a bond forms between the inhibitor Si atom and the silanol O atom (Figure~\ref{fig:mechanism}a). At a siloxane bridge, one Si--O bond of the bridge and one Si--N bond of the inhibitor break, while the silyl fragment forms a new bond with the surface oxygen and the nitrogen-containing fragment bonds through its N atom to the surface Si atom formerly bonded to that oxygen.

Figure~\ref{fig:reactivity-inh} shows $\Delta E$ and $E_\mathrm{a}$ for the inhibitors at eight silanol and five siloxane sites. To highlight the site dependence, the sites are ordered by the mean $E_\mathrm{a}$ across the three inhibitors. All reactions at silanol sites are exothermic ($\Delta E\sim -1$ eV), with only modest site-to-site variations and weak inhibitor dependence. {\sloppy Silanol reactivity is not restricted to a particular configuration: isolated (\ce{\equiv Si-OH}), vicinal (\ce{HO-Si-O-Si-OH}), and geminal (\ce{=Si-(OH)2}) groups can all participate in the reaction. In contrast, reactions at siloxane bridges are mostly exothermic but generally less energetically favorable, with endothermic reactions in some cases. 

Activation barriers at silanol sites are generally below 1~eV for DMATMS and DMADMS and slightly higher for HMDS, suggesting that adsorption is kinetically accessible at experimental temperatures. The order DMATMS $<$ DMADMS $<$ HMDS holds for $E_\mathrm{a}$ at nearly every site. This trend is consistent with differences in the proton-accepting ability and steric accessibility of the inhibitor nitrogen. The dimethylamino nitrogen in DMATMS and DMADMS readily accepts the silanol proton during the transition from R to P, whereas the nitrogen in HMDS is a weaker base, and its two trimethylsilyl groups impose additional steric hindrance. The slightly higher barriers of DMADMS compared with DMATMS would reflect additional steric hindrance from its second dimethylamino group during the reaction. The predicted barrier ordering is consistent with experimental observation that DMATMS adsorption reaches saturation faster than HMDS adsorption despite a lower inhibitor pressure~\cite{Dongen.2023}. In contrast, adsorption at siloxane sites is more kinetically challenging because of the higher activation barriers associated with cleavage of a strong \ce{Si-O} bond and substantial structural distortion. Overall, these results identify silanol groups as the primary active sites for inhibitor adsorption, consistent with experimental observations~\cite{Khan.2018,Soethoudt.2019}.\par} 

An alternative inhibitor adsorption pathway involves reaction of a methyl group with a surface silanol to release methane (\ce{CH4}), but this pathway exhibits substantially higher activation barriers (Supplementary Note~S4). These higher barriers can be rationalized by differences in proton acceptance: nitrogen can accept the proton through its lone pair, whereas methane formation requires rearrangement of the Si--C bonding electrons during proton transfer. This pathway is therefore kinetically less accessible under experimental conditions. 

The reaction energies $\Delta E$ reported here describe the UR\,$\to$\,P step and do not include entropic contributions. Including these contributions leads to a slight increase in the reaction free energy with temperature, although most reactions remain thermodynamically favorable at experimentally relevant temperatures (Table~S2). The negative $\Delta E$ values therefore reflect favorable potential-energy changes associated with bond rearrangement but do not, by themselves, establish thermodynamic favorability under experimental conditions. Subsequent byproduct desorption further lowers the free energy, making the overall UR\,$\to$\,UP reaction more favorable than the UR\,$\to$\,P step at the same temperature. Although $\Delta G$ increases with temperature for both processes, including byproduct desorption provides an additional thermodynamic driving force for adsorption. Despite the reduced thermodynamic driving force with increasing temperature, an appropriate increase in temperature is required in experiments to promote adsorption by enabling the system to overcome the activation barriers.

\subsubsection{Precursor adsorption}
\label{sec:path-surf-prec}

\begin{figure*}[t]
\centering
\includegraphics[width=\textwidth]{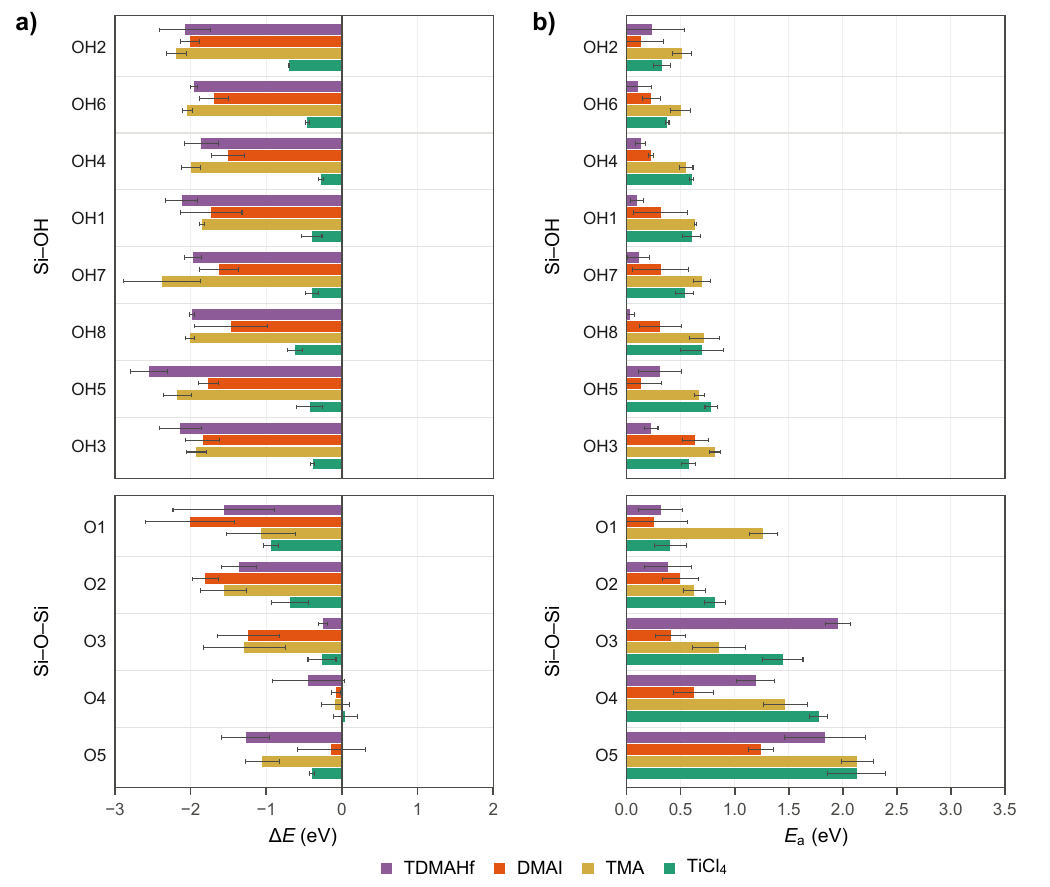}
\caption{(a) Reaction energies $\Delta E$ and (b) activation energies $E_\mathrm{a}$ of the four precursors at the sampled silanol (Si--OH) and siloxane (Si--O--Si) sites, shown using the same conventions as in Figure~\ref{fig:reactivity-inh}. Each row corresponds to a surface site numbered as in Figure~\ref{fig:sites}. Within each site type, rows are ordered from top to bottom by increasing activation energy averaged over the four precursors.}
\label{fig:reactivity-pre}
\end{figure*}

The metal precursors follow adsorption pathways distinct from those of the inhibitors. Their metal centers may accommodate an additional bond, enabling coordination to a surface oxygen to form a Lewis acid--base complex (PC and RC) before cleavage of a metal--ligand bond (Figure~\ref{fig:mechanism})~\cite{Richey.2020}. The formation of RC and PC depends on both the precursor and the surface site, and the dominant adsorption pathways for each precursor are summarized in Table~\ref{tab:pathways}. TMA follows the pathway R\,$\to$\,RC\,$\to$\,P at both silanol and siloxane sites, with the transition state occurring during the RC\,$\to$\,P step, in which an \ce{Al-CH3} bond of the precursor is cleaved and an \ce{Al-O} bond forms with surface oxygen. At silanol sites, this step also involves proton transfer to the methyl ligand, producing \ce{CH4}. \ce{TiCl4} follows the same pathway at silanol sites but reacts directly from R to P at most siloxane bridges without forming RC or PC. Adsorption forms a \ce{Ti-O} bond at both site types, while at silanol sites, proton transfer to a chloride ligand produces \ce{HCl}. DMAI and TDMAHf tend to follow the pathway R\,$\to$\,RC\,$\to$\,PC\,$\to$\,P at silanol sites, with a transition state in the RC\,$\to$\,PC step. At siloxane bridges, DMAI largely passes through both RC and PC, whereas TDMAHf forms no RC and reacts from R to PC or directly to P. The reported activation barriers correspond to the RC\,$\to$\,PC, R\,$\to$\,PC, or R\,$\to$\,P step, depending on the pathway. During DMAI adsorption, the \ce{Al-O} bond to the isopropoxide ligand is cleaved, and a new \ce{Al-O} bond forms with surface oxygen; at silanol sites, proton transfer to the isopropoxide ligand also produces isopropanol. TDMAHf adsorption involves cleavage of an \ce{Hf-N} bond of the precursor and formation of an \ce{Hf-O} bond with surface oxygen.

\begin{table}[pos=t]
\caption{Major adsorption pathways of the metal precursors at the sampled silanol (Si--OH) and siloxane bridge (Si--O--Si) sites. Double arrows indicate the steps containing the transition states whose barriers are reported as $E_\mathrm{a}$ in Figure~\ref{fig:reactivity-pre}.
}
\label{tab:pathways}
\begin{tabular*}{\tblwidth}{@{} LLLL@{} }
\toprule
 & & \multicolumn{2}{c}{Pathway} \\
\cmidrule(lr){3-4}
Precursor & X & Si--OH & Si--O--Si \\
\midrule
TMA & \ce{CH3} & R\,$\to$\,RC\,$\Rightarrow$\,P & R\,$\to$\,RC\,$\Rightarrow$\,P \\
\ce{TiCl4} & Cl & R\,$\to$\,RC\,$\Rightarrow$\,P & R\,$\Rightarrow$\,P \\
DMAI & \ce{O^{i}Pr} & R\,$\to$\,RC\,$\Rightarrow$\,PC\,$\to$\,P & R\,$\to$\,RC\,$\Rightarrow$\,PC\,$\to$\,P \\
TDMAHf & \ce{NMe2} & R\,$\to$\,RC\,$\Rightarrow$\,PC\,$\to$\,P & R\,$\Rightarrow$\,PC\,$\to$\,P \\
\bottomrule
\end{tabular*}
\end{table}

At silanol sites, all four precursors exhibit negative $\Delta E$ values, indicating energetically favorable adsorption. The reaction energies are mostly below $-1.5$~eV for TMA, DMAI, and TDMAHf, whereas those for \ce{TiCl4} are close to $-0.5$~eV. The activation barriers remain below 1~eV for all four precursors, highlighting the importance of surface \ce{-OH} groups as reactive sites for precursor adsorption. TMA slightly exhibits higher activation barriers than the other precursors despite its more favorable reaction energies. As discussed above, this behavior reflects differences in proton acceptance: methane formation involves rearrangement of the metal--carbon bonding electrons, whereas lone pairs on the proton-accepting atoms (O, N, and Cl) of the other ligands can facilitate proton transfer. For \ce{TiCl4}, chloride lone pairs can facilitate proton transfer, but its reaction energies are more than 1~eV higher than those of the other precursors, indicating substantially less favorable ligand-exchange energetics. This thermodynamic disadvantage could partly offset the favorable proton-accepting character of chloride, resulting in activation barriers comparable to those of TMA.

At siloxane bridges, precursor adsorption generally yields negative $\Delta E$ values (Figure~\ref{fig:reactivity-pre}). As with the inhibitors, activation barriers are typically higher than at silanol sites. However, some sites, such as O1 and O2, exhibit comparable barriers to those at silanol sites, suggesting that adsorption is kinetically accessible at these bridges. No systematic ordering of adsorption energetics is observed among the four precursors at siloxane sites. This may reflect the greater complexity of adsorption at siloxane sites, where substantial structural distortions contribute to the reaction energetics and can obscure trends associated with precursor chemistry.

\section{Coverage of the inhibitor layer}
\label{sec:layer}

To evaluate inhibitor coverage, which is critical for suppressing subsequent ALD growth on non-growth areas, we perform off-lattice GCMC simulations. For simplicity, only chemisorption at silanol sites is included, based on the site reactivity discussed above. The reaction energy used to evaluate the acceptance criterion is calculated on the fly using the MLIP after structural relaxation. A steric filter rejects trial configurations in which the intact inhibitor molecule overlaps with neighboring adsorbates. In addition to adsorption from the gas phase, we include a second ligand-exchange reaction for adsorbed DMADMS, which has two dimethylamino ligands, when neighboring silanol groups are available. The criteria used to identify neighboring silanol groups for this additional reaction are detailed in Supplementary Note~S5.

\begin{figure}[t]
\centering
\includegraphics[width=\columnwidth]{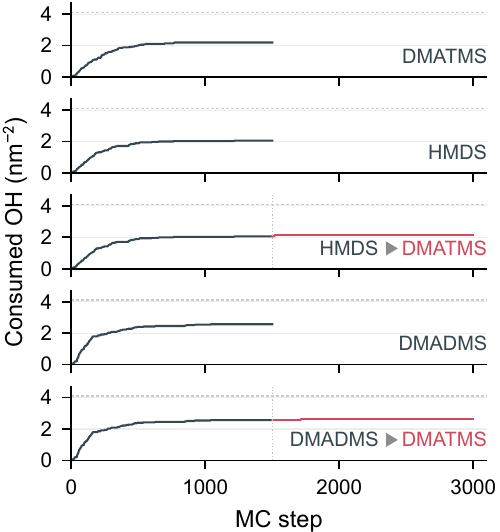}
\caption{Consumed silanol density as a function of MC step during off-lattice GCMC simulations, shown for one run of each dosing sequence: single doses of DMATMS, HMDS, and DMADMS, and sequential doses in which an HMDS or DMADMS dose is followed by a DMATMS dose. The dashed horizontal line indicates the silanol density of the bare surface (4.11~OH/nm$^2$). The vertical dotted line marks the start of the subsequent DMATMS dose, and the red curves show the consumed silanol density during this dose.}
\label{fig:convergence}
\end{figure}

As shown in Figure~\ref{fig:convergence}, the density of consumed silanol groups associated with the inhibitor coverage rises steeply during the first few hundred MC steps and reaches a stable plateau within 1500 steps. Across five independent GCMC runs, DMATMS consumes $2.23 \pm 0.05$~OH/nm$^2$ from an initial silanol density of 4.11~OH/nm$^2$, leaving about 1.9~OH/nm$^2$ unreacted. Because each DMATMS molecule consumes one silanol group, the trimethylsilyl density equals the consumed silanol density and agrees with the experimental saturation density of 2.2--2.3~nm$^{-2}$ for DMATMS on hydroxylated \ce{SiO2}~\cite{Soethoudt.2019,Dongen.2023}. The simulations also reproduce the experimental observation that some silanol groups remain unreacted at saturation because of steric hindrance~\cite{Soethoudt.2019,Clerix.2023}. DMADMS consumes the largest number of silanol groups, reaching 2.6~OH/nm$^2$. This is because its two reactive dimethylamino ligands allow some molecules to form doubly anchored surface species, each consuming two silanol groups. Approximately 30\% of the adsorbed DMADMS fragments are doubly anchored. The adsorbed fragment density itself is approximately 2.0~nm$^{-2}$, close to that of DMATMS. 

HMDS reaches saturation at a consumed silanol density 0.1--0.2~OH/nm$^2$ lower than that of DMATMS, although both inhibitors leave the same trimethylsilyl group on the surface and undergo exothermic chemisorption. In our simulations, this difference arises from the steric filter: the intact HMDS molecule, containing two trimethylsilyl groups linked by nitrogen, requires more space than DMATMS to approach an available site. As coverage increases, fewer sites remain accessible to HMDS, resulting in a slightly lower saturation coverage. 

\begin{figure*}[t]
\centering
\includegraphics[width=\textwidth]{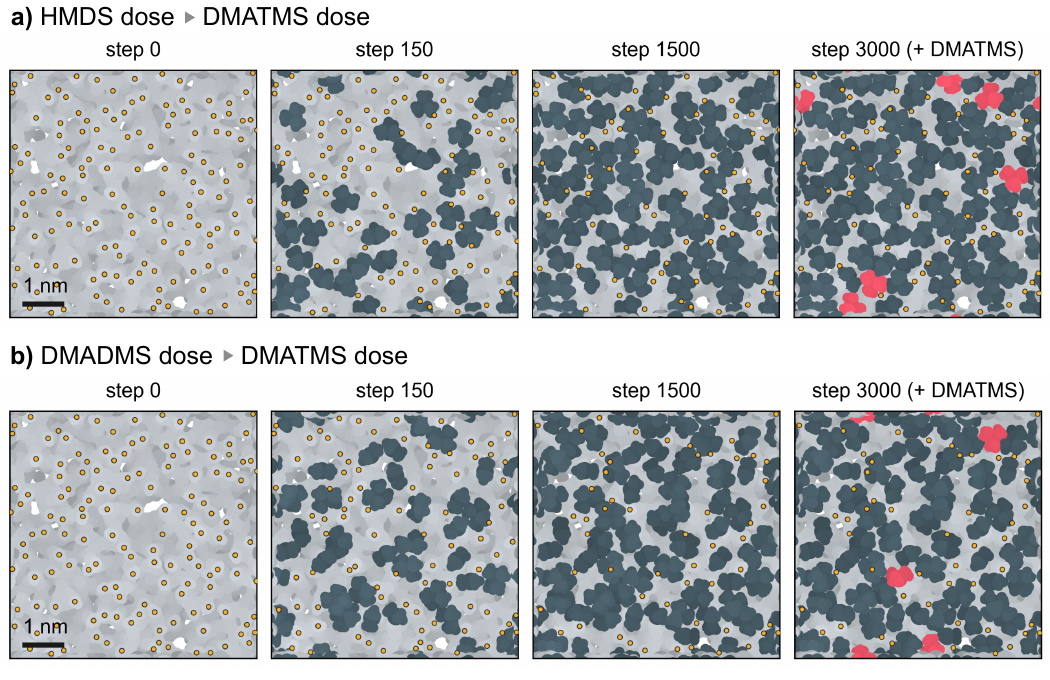}
\caption{Top views of the inhibitor layer during off-lattice GCMC simulations for (a) an HMDS dose followed by a DMATMS dose and (b) a DMADMS dose followed by a DMATMS dose. Snapshots are shown at MC steps 0, 150, and 1500 during the first dose and after 1500 additional steps of the DMATMS dose (step 3000). Atoms are represented by van der Waals disks projected onto the surface plane. Fragments chemisorbed during the first dose are shown in dark gray and those added during the subsequent DMATMS dose in red. Unreacted silanol groups are marked by orange dots, and the a-\ce{SiO2} surface is shaded gray, with lighter shades indicating greater height. The scale bar represents 1~nm.}
\label{fig:coverage}
\end{figure*}

\begin{figure}[t]
\centering
\includegraphics[width=\columnwidth]{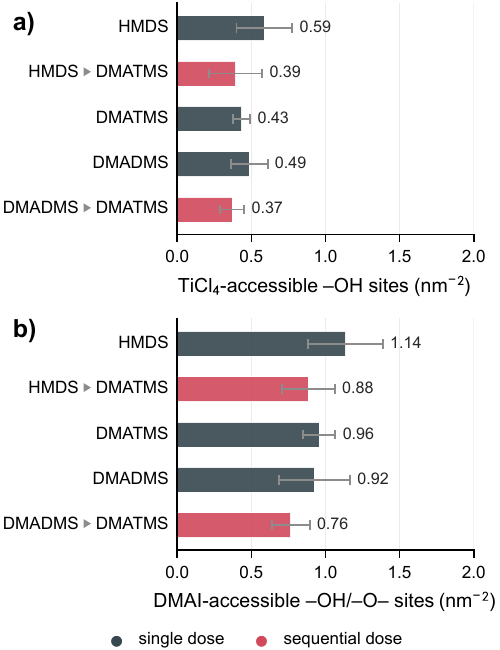}
\caption{Density of accessible sites remaining on the inhibitor-passivated surface: (a) silanol groups accessible to a \ce{TiCl4} probe and (b) silanol groups and exposed siloxane bridges accessible to a DMAI probe. Dark bars represent a single inhibitor dose, and red bars represent an additional DMATMS dose. Bars show means over five independent runs, with error bars indicating standard deviations.}
\label{fig:accessible}
\end{figure}

We next quantify the remaining reactive sites accessible to each precursor by placing a probe molecule at each site and applying the steric filter (Figure~\ref{fig:accessible}). To compare with experiments that systematically investigated inhibitor effects, we focus on \ce{TiCl4} and DMAI~\cite{Dongen.2023,Xu.202223f}. Because \ce{TiCl4} adsorption is strongly favored at silanol sites and substantially less favorable at siloxane bridges (Section~\ref{sec:path-surf-prec}), its accessibility is evaluated only at the remaining Si--OH groups. For DMAI, which can adsorb at both site types, accessibility is evaluated at the remaining Si--OH groups and exposed Si--O--Si bridges. On the bare surface, the probes can access 97--99\% of the candidate sites, providing a baseline for quantifying site blocking by the inhibitor layer. The accessible-site density for \ce{TiCl4} is found to be $0.59 \pm 0.19$, $0.43 \pm 0.06$, and $0.49 \pm 0.12$~nm$^{-2}$ for HMDS, DMATMS, and DMADMS, respectively. The corresponding densities for DMAI are $1.14 \pm 0.25$, $0.96 \pm 0.11$, and $0.92 \pm 0.24$~nm$^{-2}$.

HMDS leaves approximately 35\% more accessible sites than DMATMS for \ce{TiCl4} and 20\% more for DMAI. Although both inhibitors produce the same surface group [\ce{-(CH3)3}], the lower coverage achieved by HMDS results in less effective site blocking. This trend is consistent with the experimentally observed superior suppression of \ce{TiO2} ALD using \ce{TiCl4} on DMATMS-treated \ce{SiO2} compared with HMDS-treated \ce{SiO2}, with selectivities of 0.90 and 0.13, respectively, after 100 cycles~\cite{Dongen.2023}. Our simulations suggest that steric constraints on inhibitor uptake contribute to this difference, rather than differences in the chemistry of the anchored fragments. In addition, the higher activation barriers for HMDS adsorption further limits the coverage achieved during finite experimental exposures, requiring longer doses or higher temperatures to approach saturation and potentially reducing its blocking effectiveness under comparable dosing conditions~\cite{Dongen.2023}. DMADMS leaves approximately as many accessible sites as DMATMS for both precursors. This suggests that the greater silanol consumption by doubly anchored DMADMS fragments does not translate into a substantial improvement in site blocking. This finding is consistent with previous experiments reporting comparable blocking effectiveness for DMADMS and DMATMS treatments~\cite{Xu.202223f}.

Sequential inhibitor dosing has been employed experimentally, using HMDS followed by DMATMS for \ce{TiO2} ALD with \ce{TiCl4}~\cite{Dongen.2023} and DMADMS followed by DMATMS for \ce{Al2O3} ALD with DMAI~\cite{Xu.202223f}. We simulate both dosing sequences to evaluate their effects on the remaining precursor-accessible sites. Figure~\ref{fig:coverage} shows the evolution of inhibitor coverage during GCMC sampling. In both sequences, coverage by the first inhibitor, HMDS or DMADMS, largely saturates within 1500 MC steps (dark gray). Some silanol groups nevertheless remain unreacted because steric hindrance prevents the intact inhibitor molecules from approaching them (orange dots). During the second dose, the smaller DMATMS molecule can access and react with some of these residual silanol groups (red), increasing the surface coverage. Consequently, HMDS followed by DMATMS reduces the \ce{TiCl4}-accessible site density from 0.59 to 0.39~nm$^{-2}$, comparable to or slightly below that obtained with a single DMATMS dose (Figure~\ref{fig:accessible}). Similarly, DMADMS followed by DMATMS reduces the DMAI-accessible site density from 0.92 to 0.76~nm$^{-2}$. These results are consistent with the enhanced blocking effectiveness observed experimentally for sequential inhibitor dosing~\cite{Dongen.2023,Xu.202223f}.

\section{Outlook}
\label{sec:path-inh-prec}

So far, we have focused on inhibitor surface coverage, but blocking capability also depends on the reactivity of the adsorbed inhibitors toward precursors and co-reactants. Previous experiments reported that aminosilane-treated \ce{SiO2} surfaces effectively suppress \ce{TiCl4} adsorption without inhibitor degradation, whereas TMA partially degrades the inhibitors during precursor exposure~\cite{Soethoudt.2020}. We indeed find the distinct reaction energetics of the adsorbed inhibitor fragment with \ce{TiCl4} and TMA; we evaluate the reaction energies and activation barriers for a possible ligand-exchange pathway involving a methyl \ce{C-H} bond in the adsorbed inhibitor:
\begin{equation}
\equiv\!\mathrm{Si}{-}\mathrm{CH}_3
+ \mathrm{M}{-}\mathrm{X}
\longrightarrow
\equiv\!\mathrm{Si}{-}\mathrm{CH}_2{-}\mathrm{M}
+ \mathrm{H}{-}\mathrm{X}.
\end{equation}
For \ce{TiCl4}, this pathway is endothermic and has a high activation barrier, with $\Delta E=0.90\pm0.07$~eV and $E_{\mathrm{a}}=1.99\pm0.09$~eV (averaged over three configurations at a single site). In contrast, TMA exhibits more favorable reaction energetics and a lower activation barrier, with $\Delta E=-0.64\pm0.07$~eV and $E_{\mathrm{a}}=1.33\pm0.09$~eV averaged over five configurations at the same site. These results suggest that TMA is more likely than \ce{TiCl4} to modify the adsorbed inhibitor fragments through this pathway, consistent with the experimentally observed difference in inhibitor stability. This example illustrates the importance of investigating precursor-induced inhibitor degradation in theoretical studies of AS-ALD.

As demonstrated above, inhibitor and precursor reactions on amorphous substrates can vary significantly across reaction sites. This variability reflects differences in local bonding environments, surface functional groups, and steric accessibility, which influence both reaction energies and activation barriers. Although many previous theoretical studies have employed crystalline substrate models for computational simplicity, such models represent a limited set of well-defined surface environments and may not capture the heterogeneity of amorphous surfaces. Consequently, conclusions drawn from selected crystalline sites may differ even qualitatively from those obtained for amorphous substrates, particularly regarding the relative reactivity of inhibitors and precursors. Moreover, inhibitor coverage and blocking capability depend not only on the reactivity of individual sites but also on their spatial distribution and the accessibility of neighboring sites. Realistic substrate models and adequate sampling of diverse local environments are therefore essential for investigating adsorption mechanisms and predicting inhibition performance in AS-ALD.

\section{Conclusions}
\label{sec:conclusions}

In this work, we presented a computational workflow combining a pretrained universal MLIP with off-lattice GCMC simulations and applied it to investigate inhibitor adsorption and precursor blocking on amorphous \ce{SiO2}. We constructed realistic a-\ce{SiO2} surface models representative of experimental conditions. The calculations identified silanol groups as the primary adsorption sites for aminosilane and related inhibitors and various metal precursors, revealing substantial variations in reaction energetics across local surface environments. Coverage simulations based on GCMC further showed that inhibitors yielding the same adsorbed fragment can achieve different coverages because the intact molecules impose distinct steric constraints, resulting in different accessible site densities and blocking capabilities. We also highlighted the importance of inhibitor-fragment reactivity toward precursors and co-reactants, realistic amorphous substrate models, and adequate sampling of diverse reaction sites. The proposed workflow and chemical insights gained from these case studies provide guidance for future computational studies of AS-ALD and the selection of suitable inhibitors and precursors.

\section*{Declaration of competing interest}

The authors declare that they have no known competing financial interests or personal relationships that could have appeared to influence the work reported in this paper.

\section*{Acknowledgements}

The computations were carried out at Korea Institute of Science and Technology Information (KISTI) National Supercomputing Center (KSC-2025-CRE-0284). This work was supported by Samsung Electronics South Korea  under grant code IO251222-14917-01 and the Nano \& Material Technology Development Program through the National Research Foundation of Korea(NRF) funded by Ministry of Science and ICT(RS-2026-25586219). 

\appendix

\printcredits

\bibliographystyle{model1-num-names}

\bibliography{cas-refs}

\end{document}


\maketitle

\section{Structure of the bulk amorphous \ce{SiO2}}
\label{si:bulk}
\begin{figure}[!htb]
\centering
\includegraphics{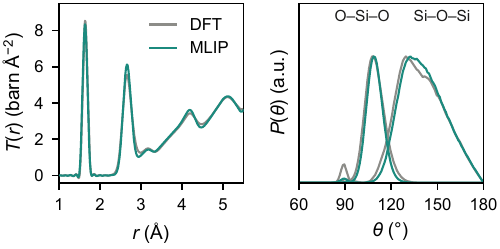}
\caption{Structure of the bulk amorphous \ce{SiO2}. Left: neutron-weighted total correlation function $T(r)$ of the 2592-atom MLIP cells (teal) and of the 96-atom DFT melt--quench structures of Kim et al.~\cite{Kim.2026pecvd} (gray). Right: bond-angle distributions $P(\theta)$ of O--Si--O and Si--O--Si for the same structures, each normalized to its maximum.}
\label{fig:si-peaks}
\end{figure}

The melt--quench cells are checked against DFT before cleaving. Figure~\ref{fig:si-peaks} compares the neutron-weighted total correlation function $T(r)$ and the O--Si--O and Si--O--Si bond-angle distributions of the three 2592-atom MLIP cells with those of the DFT melt--quench structures of Kim et al.~\cite{Kim.2026pecvd}, ten 96-atom cells at the same density of 2.2~g/cm$^3$. The MLIP cells reproduce the peak positions and heights of $T(r)$ and both bond-angle distributions of the DFT structures.

\section{Surface models and silanol density}
\label{si:dehydroxylation}
Under-coordinated Si and O atoms on the cleaved faces are identified by comparing the coordination number of every atom before and after the cut. An oxygen that loses two or more bonds is moved to the opposite face to recover its original coordination before the remaining dangling bonds are terminated.

Following Ewing et al.~\cite{Ewing.2014}, the surface is dehydroxylated by condensing neighboring silanol pairs one at a time, each condensation releasing one water molecule. The free-energy change of a condensation is
\begin{equation}
\Delta G(T,P) = E_\text{after} - E_\text{before} + \mu_{\ce{H2O}}(T,P) ,
\label{eq:si-condensation}
\end{equation}
where $E_\text{before}$ and $E_\text{after}$ are the relaxed energies of the slab before and after the condensation and $\mu_{\ce{H2O}}(T,P)$ is the chemical potential of gas-phase water at temperature $T$ and water partial pressure $P$. A condensation is thus favorable, $\Delta G(T,P) < 0$, when its condensation energy $\Delta E = E_\text{after} - E_\text{before}$ is below $-\mu_{\ce{H2O}}(T,P)$.

The chemical potential of a gas-phase molecule $X$ is evaluated as
\begin{equation}
\mu_X(T,P) = E_X - k_\mathrm{B} T \ln\left[q^{\text{tr}}_X(T,P_0)\, q^{\text{rot}}_X(T)\right] + k_\mathrm{B} T \ln\frac{P}{P_0},
\label{eq:si-mux}
\end{equation}
from the energy $E_X$ of the isolated molecule on the same potential and the standard ideal-gas translational and rotational partition functions,
\begin{equation}
q^{\text{tr}}_X(T,P_0) = \left(\frac{2\pi m k_\mathrm{B} T}{h^2}\right)^{3/2}\frac{k_\mathrm{B} T}{P_0},
\end{equation}
\begin{equation}
q^{\text{rot}}_X(T) = \frac{\sqrt{\pi}}{\sigma}\left(\frac{8\pi^2 k_\mathrm{B} T}{h^2}\right)^{3/2}\sqrt{I_A I_B I_C},
\end{equation}
where $m$ is the mass of the molecule, $\sigma$ its rotational symmetry number ($\sigma = 2$ for water), and $I_A$, $I_B$, and $I_C$ its principal moments of inertia. $q^{\text{tr}}_X$ is the partition function per molecule at the reference pressure $P_0$. Vibrational contributions and zero-point energies are omitted on the assumption, taken from Ewing et al.~\cite{Ewing.2014}, that they cancel between the gas molecules and the surface.

On the 2592-atom surfaces the condensation energies of the candidate pairs are cached between successive condensations. Only the energies of pairs within 10~\AA{} of the last condensed pair are recomputed. The lowest-$\Delta E$ candidate and every candidate within 0.1~eV of it are recomputed before a condensation is accepted, and all cached values are recomputed before the dehydroxylation stops. 

\begin{figure}[!htb]
\centering
\includegraphics{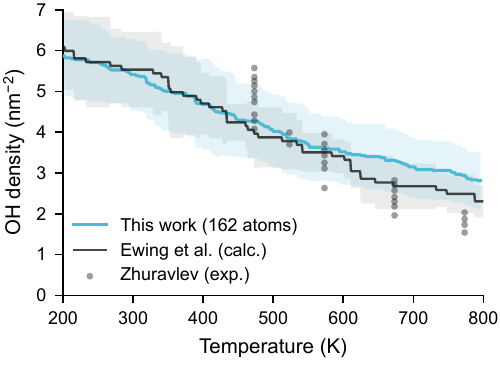}
\caption{Silanol density of the 162-atom surfaces as a function of temperature at a water partial pressure of $10^{-6}$~bar. The line represents the average over 40 surfaces (both faces of 20 slabs), and the shaded band indicates their variation. The calculated values of Ewing et al.~\cite{Ewing.2014} and the experimental values of Zhuravlev~\cite{Zhuravlev.2000}, digitized from Figure~9 of that work, are shown for comparison.}
\label{fig:si-oh-sizes}
\end{figure}

Figure~\ref{fig:si-oh-sizes} shows the silanol density of the 162-atom surfaces used for the reactivity calculations. They follow the same trend with temperature as the 2592-atom surfaces in the main text.

\section{Validation of the MLIP and the cell size}
\label{si:validation}
\begin{figure}[!htb]
\centering
\includegraphics{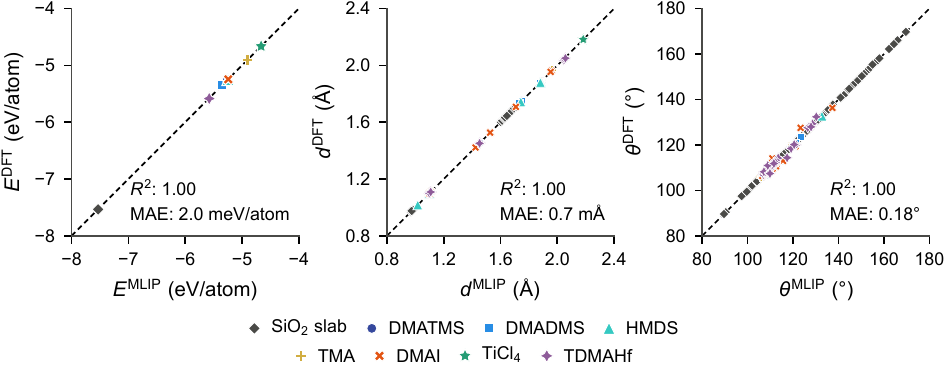}
\caption{Total energies per atom (left), bond lengths (middle), and bond angles (right) of the bare \ce{SiO2} slab and the seven isolated molecules relaxed with the MLIP and with DFT. Symbols and colors identify the structures, with the molecules colored as in the main text. The dashed line is the parity line, and $R^2$ of the linear fit and the mean absolute error are given in each panel.}
\label{fig:si-energy}
\end{figure}

The reference states are the bare slab of the reactivity calculations and the seven molecules, each in a 20~\AA{} box. Each structure is relaxed again with both potentials, starting from the same MLIP minimum. The atoms in the bottom region of the slab are held fixed. Figure~\ref{fig:si-energy} compares the total energies per atom, the bond lengths, and the bond angles of the two sets of minima, where for the slab only the bonds and angles that involve at least one free atom are counted. The energies agree to 2.0~meV/atom, the bond lengths to 0.7~m\AA{}, and the bond angles to 0.18$^\circ$ on average, so the MLIP reproduces both the energies and the structures of the reference states.

\begin{figure}[!htb]
\centering
\includegraphics{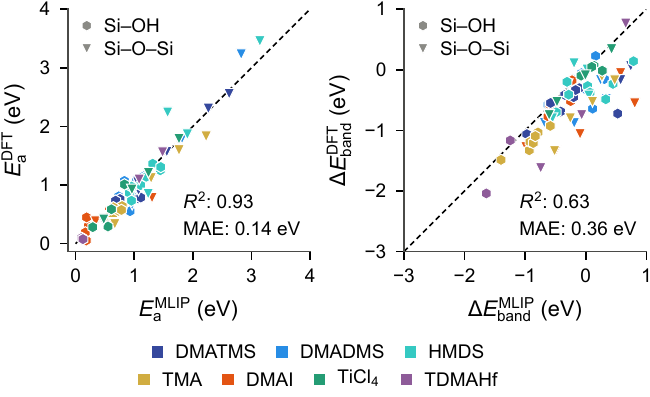}
\caption{Activation energies (left) and band reaction energies $\Delta E_\mathrm{band}$ (right) of the 72 validated reactions computed with the MLIP and with DFT. Circles mark reactions at silanol (Si--OH) sites and triangles reactions at siloxane (Si--O--Si) sites. Colors identify the molecules as in the main text. The dashed line is the parity line, and $R^2$ of the linear fit and the mean absolute error are given in each panel. $\Delta E_\mathrm{band}$ is the energy of the last image of the band relative to its first on the same potential-energy surface and differs from the $\Delta E$ of the main text, which is referenced to the separated reactants.}
\label{fig:si-parity}
\end{figure}

To verify the pretrained MLIP, we randomly select one reaction from the computed paths for each molecule--site pair in the main text and recompute it with DFT. The reactant and product are relaxed on the DFT potential-energy surface, and the path between these two minima is then recomputed with the climbing-image nudged elastic band method (CI-NEB)~\cite{Henkelman.2000} on the same surface, so that each potential is compared on its own stable states. Most of these calculations converge, and Figure~\ref{fig:si-parity} compares the two potentials for the 72 converged reactions, which cover all seven molecules and all thirteen sites. The activation energies agree to a mean absolute error of 0.14~eV with no systematic offset, so the barriers on which the comparison in the main text rests are reliable. The reaction energy $\Delta E_\mathrm{band}$ compared in the figure is the energy of the last image of the band relative to its first. It differs from the $\Delta E$ of the main text, which is referenced to the separated reactants. $\Delta E_\mathrm{band}$ shows a slightly larger deviation of 0.36~eV on average. This deviation arises mainly from the orientation of the physisorbed byproduct and from dative bonding in the product. The physisorbed byproduct does not enter the grand canonical Monte Carlo (GCMC) simulations, and the DFT values are systematically lower than the MLIP values, so the conclusions of the main text are unaffected.

The dependence of the reaction energy on the lateral cell size is tested on a (001) slab of $\alpha$-quartz with both faces covered by geminal silanol groups. The slab thickness, the vacuum layer, and the fixed bottom atoms are set as for the a-\ce{SiO2} slabs. Its silanol density of 9.5~OH/nm$^2$ is far above that of the a-\ce{SiO2} surfaces. Several molecular orientations of DMATMS are compared, and the one with the most negative $\Delta E$ is used for all cell sizes (Table~\ref{tab:si-cellsize}). The 162-atom a-\ce{SiO2} slab used for the reactivity calculations has a lateral length of 14.8~\AA{} and an area of 219.0~\AA$^2$, between those of the $3\times3$ and $4\times4$ cells.

\begin{table}[!htb]
\centering
\caption{Reaction energy $\Delta E$ of DMATMS on the (001) $\alpha$-quartz slab as a function of the lateral cell size, and its deviation from the $6\times6$ value.}
\label{tab:si-cellsize}
\vspace{2mm}
\begin{tabular}{lrrrr}
\hline
Cell & $a = b$ (\AA) & Area (\AA$^2$) & $\Delta E$ (eV) & $\Delta E - \Delta E_{6\times6}$ (meV) \\
\hline
$3\times3$ & 14.77 & 188.9 & $-1.372$ & $+46$ \\
$4\times4$ & 19.69 & 335.9 & $-1.396$ & $+22$ \\
$5\times5$ & 24.62 & 524.8 & $-1.414$ & $+4$ \\
$6\times6$ & 29.54 & 755.7 & $-1.418$ & $0$ \\
\hline
\end{tabular}
\end{table}

\section{Reaction free energies and the methane pathway}
\label{si:pathways}

Table~\ref{tab:si-dG} lists the Gibbs free energies of reaction discussed in the main text, for one representative reaction of each molecule.

\begin{table}[!htb]
\centering
\caption{Reaction energies $\Delta E$ and Gibbs free energies of reaction $\Delta G$ (eV) for one representative reaction of each molecule at the silanol OH6 (numbered as in the main text), at a partial pressure of 5~Torr for every gas-phase species. UR is the clean slab with the molecule in the gas phase. P is the chemisorbed product with the byproduct retained in the simulation cell, or for DMAI and TDMAHf the product complex PC with the byproduct coordinated to the metal. UP is the surface product relaxed again after removal of the byproduct, with the byproduct in the gas phase. The byproduct is given in parentheses.}
\label{tab:si-dG}
\vspace{2mm}
\begin{tabular}{lrrrrrr}
\hline
 & $\Delta E$ & \multicolumn{5}{c}{$\Delta G$} \\
\cline{3-7}
 & & 100~K & 200~K & 300~K & 400~K & 500~K \\
\hline
\multicolumn{7}{l}{UR\,$\to$\,P, byproduct retained in the cell} \\
DMATMS & $-1.35$ & $-1.02$ & $-0.79$ & $-0.55$ & $-0.32$ & $-0.09$ \\
DMADMS & $-1.26$ & $-0.89$ & $-0.62$ & $-0.36$ & $-0.10$ & $+0.16$ \\
HMDS & $-1.88$ & $-1.51$ & $-1.23$ & $-0.96$ & $-0.68$ & $-0.40$ \\
TMA & $-2.05$ & $-1.76$ & $-1.55$ & $-1.35$ & $-1.15$ & $-0.95$ \\
DMAI (PC) & $-1.85$ & $-1.49$ & $-1.21$ & $-0.92$ & $-0.64$ & $-0.36$ \\
\ce{TiCl4} & $-0.54$ & $-0.36$ & $-0.12$ & $+0.11$ & $+0.33$ & $+0.54$ \\
TDMAHf (PC) & $-2.49$ & $-2.01$ & $-1.69$ & $-1.37$ & $-1.04$ & $-0.72$ \\
\hline
\multicolumn{7}{l}{UR\,$\to$\,UP, byproduct in the gas phase} \\
DMATMS (\ce{HNMe2}) & $-0.95$ & $-0.86$ & $-0.82$ & $-0.78$ & $-0.74$ & $-0.69$ \\
DMADMS (\ce{HNMe2}) & $-0.96$ & $-0.83$ & $-0.77$ & $-0.70$ & $-0.63$ & $-0.56$ \\
HMDS (\ce{Me3SiNH2}) & $-0.82$ & $-0.77$ & $-0.74$ & $-0.70$ & $-0.67$ & $-0.64$ \\
TMA (\ce{CH4}) & $-1.91$ & $-1.78$ & $-1.70$ & $-1.63$ & $-1.55$ & $-1.47$ \\
DMAI (\ce{^{i}PrOH}) & $-0.49$ & $-0.41$ & $-0.37$ & $-0.33$ & $-0.28$ & $-0.23$ \\
\ce{TiCl4} (HCl) & $-0.14$ & $-0.17$ & $-0.11$ & $-0.06$ & $-0.01$ & $+0.03$ \\
TDMAHf (\ce{HNMe2}) & $-1.60$ & $-1.44$ & $-1.36$ & $-1.28$ & $-1.20$ & $-1.12$ \\
\hline
\end{tabular}
\end{table}

Besides the \ce{Si-N} exchange, we compute the exchange through a \ce{Si-CH3} bond, which releases methane as the Al--CH$_3$ ligands of TMA do, for DMATMS at the silanol OH4:
\begin{align}
  \ce{$\equiv$Si-OH + Me3Si-NMe2 &-> $\equiv$Si-O-SiMe3 + HNMe2}
  \label{eq:si-n}\\
  \ce{$\equiv$Si-OH + Me3Si-NMe2 &-> $\equiv$Si-O-SiMe2NMe2 + CH4}
  \label{eq:si-c}
\end{align}
The results are listed in Table~\ref{tab:si-c}. The two exchanges are similarly exothermic, but the barrier of \ce{Si-CH3} cleavage is about 1.4~eV higher, so it is unlikely to occur. This agrees with experiment, where DMATMS releases dimethylamine and leaves a stable trimethylsilyl termination~\cite{Clerix.2023,Soethoudt.2020}.

\begin{table}[!htb]
\centering
\caption{Reaction energies $\Delta E$ and activation energies $E_\mathrm{a}$ (eV) for the two ligand-exchange reactions of DMATMS at the silanol OH4 (numbered as in the main text). Values are the mean and standard deviation over the computed paths, five for \ce{Si-N} cleavage and three for \ce{Si-CH3} cleavage. The byproduct is given in parentheses.}
\label{tab:si-c}
\vspace{2mm}
\begin{tabular}{lrr}
\hline
 & $\Delta E$ & $E_\mathrm{a}$ \\
\hline
\ce{Si-N}, Eq.~\ref{eq:si-n} (\ce{HNMe2}) & $-0.98 \pm 0.16$ & $0.76 \pm 0.10$ \\
\ce{Si-CH3}, Eq.~\ref{eq:si-c} (\ce{CH4}) & $-1.00 \pm 0.18$ & $2.15 \pm 0.19$ \\
\hline
\end{tabular}
\end{table}

\section{Off-lattice GCMC}
\label{si:gcmc}
Each step of the simulation attempts a chemical adsorption, a reverse reaction, or an exchange move, with probabilities 0.4, 0.4, and 0.2. In a chemical adsorption, the forward reaction that corresponds to an insertion in a conventional GCMC simulation, the hydrogen of the chosen silanol is removed and the inhibitor fragment is bonded through its silicon to the bared oxygen. In the reverse reaction, which corresponds to a deletion, the fragment is removed and its oxygen is recapped with hydrogen. An exchange move transfers an adsorbed fragment to another silanol at a constant number of adsorbates. After every move the structure is relaxed and the silanol sites are re-detected from the relaxed geometry. The move is then accepted or rejected with the Metropolis probabilities of Eqs.~(\ref{eq:si-accept}) and (\ref{eq:si-exch}), computed from the energy difference between the relaxed structures. The byproduct is dimethylamine for DMATMS and DMADMS, one molecule for a singly anchored (monodentate) and two for a doubly anchored (bidentate) chemical adsorption, and trimethylsilylamine for HMDS.

The simulation samples configurations $x$ of adsorbed fragments on the available sites, each with the energy $U(x)$ of its relaxed structure. A free molecule in a continuum grand canonical simulation carries the statistical weight $1/\Lambda^3$, with $\Lambda$ the thermal de Broglie wavelength, whereas an adsorbed fragment is bonded to a silanol oxygen and has no translational freedom, so its positions are the discrete sites. Its vibrations, including the frustrated translations and rotations, are neglected as in Supplementary Note~\ref{si:dehydroxylation}, so that its statistical weight is unity while the translational and rotational partition functions of the gas-phase molecules remain in the chemical potentials. In equilibrium with the gas phase, a configuration with $N_s(x)$ fragments of species $s$ then has the probability
\begin{equation}
\pi(x) \propto \exp\!\left[-\frac{U(x) - \sum_s N_s(x)\,\Delta\mu_s}{k_\mathrm{B} T}\right],
\label{eq:si-pi}
\end{equation}
where
\begin{equation}
\Delta\mu_s = \mu_\text{inh}(T,P_\text{inh}) - n_s\,\mu_\text{byp}(T,P_\text{byp})
\label{eq:si-mu}
\end{equation}
is the change in the chemical potential of the gas when one inhibitor adsorbs as species $s$ and releases $n_s$ byproduct molecules, each $\mu_X$ being evaluated with Eq.~(\ref{eq:si-mux}).

A chemical adsorption of species $s$, chosen with probability $p_s$, is proposed at one of the $M_s$ available sites with equal probability, and a reverse reaction removes one of the $N$ adsorbed fragments with equal probability. For DMADMS the monodentate and bidentate forms are proposed with equal probability, $p_s = 1/2$. The Metropolis probabilities that satisfy detailed balance with these proposals are
\begin{equation}
P_\text{ads} = \min\left\{1,\; \frac{M_s}{(N+1)\,p_s}\, \exp\!\left[-\frac{\Delta U - \Delta\mu_s}{k_\mathrm{B} T}\right]\right\}, \quad P_\text{rev} = \min\left\{1,\; \frac{N\,p_s}{M_s'}\, \exp\!\left[-\frac{\Delta U + \Delta\mu_s}{k_\mathrm{B} T}\right]\right\},
\label{eq:si-accept}
\end{equation}
where $\Delta U$ is the energy change of the proposed move, $N$ is the number of adsorbates before the move, and $M_s$ and $M_s'$ are the numbers of available sites of species $s$ before and after the move, so that $M_s'$ of a reverse reaction includes the site it frees. The factor $M_s/(N+1)$ comes from the proposal probabilities, as $V/(N+1)$ does in a continuum simulation. An exchange move conserves $N$ and $\Delta\mu_s$ and is accepted with
\begin{equation}
P_\text{ex} = \min\left\{1,\; \frac{M_s}{M_s'}\, \exp\!\left[-\frac{\Delta U}{k_\mathrm{B} T}\right]\right\}.
\label{eq:si-exch}
\end{equation}
Because every energy is that of a relaxed structure, the sampled configurations are local minima, and the placement and relaxation make the proposals reversible only approximately.

Sites blocked by the steric filter are counted in $M_s$ and $M_s'$, and the moves into or out of them are rejected. A bidentate adsorption of DMADMS requires two silanols on different silicon atoms with an O--O distance between 2.4 and 3.4~\AA. For this species, $M_s$ and $M_s'$ count bridgeable silanol pairs rather than individual silanols. A trial whose relaxed structure leaves the silicon of a bidentate fragment more than 2.1~\AA{} from either of its oxygens is rejected outright.

Molecules and fragments are placed as rigid bodies over the fixed substrate with the packing algorithm of PACKMOL~\cite{Martinez.2009} in three steps of the simulation.
\begin{enumerate}[label=(\roman*), leftmargin=*, itemsep=2pt, topsep=4pt]
\item To generate the chemisorbed structure of a chemical adsorption, the inhibitor fragment is placed with its anchor silicon at bonding distance from the target silanol oxygen, or from both target oxygens for a bidentate fragment, without overlap with the surrounding surface atoms.
\item In the steric filter applied before a chemical adsorption, the intact molecule is placed over the surface with its amine nitrogen hydrogen-bonded to the target silanol and its reactive silicon near the silanol oxygen. The adsorption is rejected if no such placement keeps every atom of the molecule at least the sum of their van der Waals radii~\cite{Alvarez.2013} away from every surface atom. The silanol hydrogen and oxygen are exempt. For a molecule with more than one amine nitrogen, a placement through any of them passes the filter.
\item To count the sites accessible to a precursor, a \ce{TiCl4} or DMAI probe is placed with its metal atom near the site oxygen under the same van der Waals criterion, and the site is counted as accessible if such a placement exists.
\end{enumerate}

\FloatBarrier
\bibliographystyle{model1-num-names}
\bibliography{cas-refs}